\documentclass[aps,prd,preprint]{revtex4-1}
\usepackage {graphicx}
\usepackage[toc,page]{appendix}
\usepackage{amsmath}
\usepackage{tikz}
\newcommand{\be}{\begin{equation}}
\newcommand{\ee}{\end{equation}}
\newcommand{\bea}{\begin{eqnarray}}
\newcommand{\eea}{\end{eqnarray}}
\begin{document}
\title{Relative velocity in special relativity and quantum field theory}
\author{David Garfinkle}
\email{garfinkl@oakland.edu}
\affiliation{Dept. of Physics, Oakland University, Rochester, MI 48309, USA}
\affiliation{Leinweber Center for Theoretical Physics, Randall Laboratory of Physics, University of Michigan, Ann Arbor, MI 48109-1120, USA}

\date{\today}

\begin{abstract}

A derivation of the relative velocity used in the definition of the relativistic cross-section is given in terms of manifestly Lorentz invariant quantities.  Along the way we find that there is a certain arbitrariness in the usual definition of cross-section.

\end{abstract}


\maketitle

In the lab frame objects 1 and 2 have velocities ${\vec v}_1$ and ${\vec v}_2$.  What is the speed $v_{\rm rel}$ of object 2 as measured by object 1?  In Newtonian mechanics, the answer is simple: $ {v_{\rm rel}}=|{{\vec v}_2}-{{\vec v}_1}|$.  However, in special relativity we get a more complicated answer.  The usual textbook approach is to consider only the case where ${\vec v}_1$ and ${\vec v}_2$ are collinear and then to make use of the Lorentz transformation connecting the lab frame to the rest frame of object 1.  This results in the following formula
\be
{v_{\rm rel}} = {\frac {|{v_2} \mp {v_1}|} {1 \mp {v_1}{v_2}}}
\label{vrel0}
\ee
Here the minus sign is for when the two objects move in the same direction while the plus sign is for when they move in opposite directions. For convenience, we use units where the speed of light $c$ is equal to one.

It is straightforward to generalize this formula to the non-collinear case by using the inner product of four-velocities.  Let the four-velocities of the two objects be respectively $u^\alpha _1$ and $u^\alpha _2$.  In the rest frame of object 1 we have ${u^\alpha _1}=(1,0,0,0)$ and ${u^\alpha _2}=\gamma (1,{\vec v})$ where $\vec v$ is the relative velocity ({\it i.e.} the velocity of object 2 as measured by object 1) and $\gamma = 1/{\sqrt {1-{v^2}}}$.  We then find
${u^\alpha _1}{u_{2\alpha}} = - \gamma = -1/{\sqrt {1-{v^2}}}$ and therefore the relative speed between the two objects is given by
\be
{v_{\rm rel}} = {\sqrt {1 - {{({u^\alpha _1}{u_{2\alpha}})}^{-2}}}}
\label{vrel1}
\ee
Since inner products are invariant, it follows that eqn. (\ref{vrel1}) is the relative speed between the two objects, no matter what reference frame we use.  Furthermore, it is straightforward to show that in the case where the velocities are collinear, eqn. (\ref{vrel1}) reduces to eqn. (\ref{vrel0}) as follows: in an arbitrary frame we have ${u^\alpha _1}={\gamma _1}(1,{{\vec v}_1})$ and ${u^\alpha _2}={\gamma _2}(1,{{\vec v}_2})$, so we have
${u^\alpha _1}{u_{\alpha 2}}= - {\gamma_1}{\gamma _2} (1- {{\vec v}_1}\cdot {{\vec v}_2 })$. It then follows from eqn. (\ref{vrel1}) that
\bea
{v_{\rm rel}} &=& {\frac  {\sqrt {{{(1 - {{\vec v}_1} \cdot {{\vec v}_2})}^2} - (1-{v_1 ^2})(1-{v_2 ^2})}} {1 - {{\vec v}_1} \cdot {{\vec v}_2}}}
\nonumber
\\ 
&=&  {\frac {\sqrt {[1-2{{\vec v}_1} \cdot {{\vec v}_2} + {{({{\vec v}_1} \cdot {{\vec v}_2})}^2}] - [1 - {v_1 ^2} - {v_2 ^2} + {v_1 ^2}{v_2 ^2} ] }} {1 - {{\vec v}_1} \cdot {{\vec v}_2}}}
\nonumber
\\
&=& {\frac {\sqrt {{{|{{\vec v}_2}-{{\vec v}_1}|}^2} - {{|{{\vec v}_2}\times {{\vec v}_1}|}^2}}}
{1 - {{\vec v}_1} \cdot {{\vec v}_2}}}
\label{vrel2}
\eea
It then follows immediately that if ${\vec v}_1$ and ${\vec v}_2$ are collinear then $v_{\rm rel}$ is given by eqn. (\ref{vrel0}).

All this is well known, so it is somewhat surprising that the quantum field theory formula for the cross-section makes use of a ``relative velocity'' (in the case where the velocities are collinear) that is simply given by $|{{\vec v}_2}-{{\vec v}_1}|$.  Furthermore, in the physically relevant case of collider physics, ${\vec v}_1$ and ${\vec v}_2$ have magnitudes only slightly less than $c$ and are oppositely directed, giving rise to a ``relative velocity'' approximately twice the speed of light!  

It is of interest to know how this seemingly anti-relativistic result comes about from a relativistically invariant theory.  To do this we will adopt an approach that uses manifestly relativistically invariant quantities throughout.  First note that in order to have orthonormal quantum field modes, the calculations of quantum field theory are done in a ``box'' of volume $V$ with periodic boundary conditions.\cite{MandlShaw}  Or, in the language of general relativity, the calculations of quantum field theory are done not on Minkowski spacetime, but rather on ${T^3} \times R$ with a flat metric.  In the end, one is supposed to take the limit as $V \to \infty$ to obtain results that apply directly to Minkowski spacetime, but this procedure means that intermediate steps in the calculation may contain powers of $V$ that will only go away in the final result. 

Consider the scattering of two particles.  The first step in the calculation is to find the rate.  This is done by choosing a particular rest frame and using the quantum Hamiltonian appropriate to that rest frame to calculate the rate at which the two particles scatter.  Put this way, neither the procedure (choice of a particular rest frame) nor the result (a rate) sound relativistically invariant.  However, having calculated a rate, one can then divide by the volume $V$ of the box to obtain a number of scattering events per unit spacetime volume, which $\emph {is}$ a relativistically invariant quantity.  

Though number of scattering events per unit spacetime volume is a relativistic invariant, it is not the final step of the calculation, as can be seen by noticing that it still contains powers of $V$.  Furthermore, the reason for these powers of $V$ is clear: since the normalization of particle states is that there is one particle in the box, large $V$ means the particles are unlikely to be sufficiently close to each other to scatter.  Thus, what we want to calculate is not a rate per unit volume, but rather a sort of ``intrinsic reactivity'' that is the rate per unit volume divided by the number densities of the particles. 

In Landau and Lifschitz vol. 2, section 12\cite{LL}, they state (based on work of Pauli\cite{Pauli}.  See also \cite{Moller}) that the cross-section is obtained by dividing the rate per unit volume by ${n_1}{n_2} {v_{\rm rel}}$.  Here $v_{\rm rel}$ is given by eqn. (\ref{vrel2}).  But what is meant by ${n_1}{n_2}$? The trouble with this phrasing is that number density is not a relativistic invariant: particles have number currents $n^\alpha$ not scalar number densities $n$.  Furthermore, the particle states used in scattering are momentum eigenstates, so for each state the number current is proportional to the four-momentum $p^\alpha$.  For massive particles, this issue is not an obstacle for the following reason: a massive particle has a unit four-velocity $u^\alpha$ and the number current points in the direction of the four-velocity, so there is a scalar $n$ such that ${n^\alpha} = n {u^\alpha}$.  Or to put it the other way around, for a massive particle one can define $n$ by $n={\sqrt {-{n^\alpha}{n_\alpha}}}$. This doesn't work for a massless particle: since its number current is null, we have ${n^\alpha}{n_\alpha}=0$.  Put another way, one invariant quantity that we might mean by ${n_1}{n_2}$ is 
${\sqrt {{n_1 ^\alpha}{n_{1\alpha}}{n_2 ^\beta}{n_{2\beta}}}}$.  
But this invariant is not suitable for defining the cross-section since it vanishes when either particle 1 or particle 2 is massless.  Furthermore, even in the case of only massive particles, this invariant is not what Landau and Lifschitz (and Pauli) have in mind for the following reason: in this case $n_1$ and $n_2$ are respectively the number densities as seen in their own rest frames and therefore do not take into account the enhancement in density that comes from Lorentz contraction.  

However, there is another invariant, which {\emph is} what Landau and Lifschitz (and Pauli) have in mind as the relativistic version of ${n_1}{n_2}$.  This invariant is
$ - {n_1 ^\alpha}{n_{2\alpha}}$.\cite{Furman,Cannoni}
We now find an expression for $ - {n_1 ^\alpha}{n_{2\alpha}}$ in terms of the particle velocities.  Since $n^\alpha$ must point in the direction of $p^\alpha$ and give rise to one particle in the box of volume $V$, it follows that 
\be
{n^\alpha} = {\frac {p^\alpha} {E V}} = {V^{-1}} (1,{\vec v})
\ee
It then follows that 
\be
- {n_1 ^\alpha}{n_{2\alpha}} = {V^{-2}} (1 - {{\vec v}_1} \cdot {{\vec v}_2})
\ee
Then from eqn. (\ref{vrel2}) we find 
\be
- {n_1 ^\alpha}{n_{2\alpha}}{v_{\rm rel}} = {V^{-2}}{\sqrt {{{|{{\vec v}_2}-{{\vec v}_1}|}^2} - {{|{{\vec v}_2}\times {{\vec v}_1}|}^2}}}
\ee
Thus for the case where ${\vec v}_1$ and ${\vec v}_2$ are collinear we obtain
\be
- {n_1 ^\alpha}{n_{2\alpha}}{v_{\rm rel}} = {V^{-2}} |{{\vec v}_2} - {{\vec v}_1}|
\ee

We now consider the role of the collinearity condition in the usual textbook statements about cross-section and relative velocity.  At first collinearity sounds like a (possibly very restrictive) statement about the particles that are scattering.  But it is really a statement about reference frames.  The four-momenta $p^\alpha _1$ and $p^\alpha _2$ span a timelike two-plane.  Any observer whose four-velocity is in that plane will see the particle velocities as collinear.  Thus any two velocities are collinear, as seen by the appropriate observer. Thus the phrase ``for velocities that are collinear'' should be read as ``as calculated by an observer for whom the two particle velocities are collinear.''  

Finally we argue that the inclusion of the factor of $v_{\rm rel}$ in the definition of the cross-section is a somewhat arbitrary choice, and that very little would change if it were omitted.  One obtains an ''intrinsic reactivity'' by dividing the rate per unit volume by the product of the number densities.  Put this way, the units of intrinsic reactivity are volume divided by time.  In the usual definition of cross-section one further divides this intrinsic reactivity by a velocity to obtain a quantity with units of area.  This allows one a nice, simple, intuitive (but also highly misleading) picture of cross-section involving hard surfaces of definite size.  It is worth considering just how little of quantum field theory would change if we were to define the cross-section by dividing by $-{n^\alpha _1}{n_{\alpha 2}}$ rather than $-{n^\alpha _1}{n_{\alpha 2}}{v_{\rm rel}}$.  The Feynman rules would be entirely unchanged since they have to do with calculating amplitudes that in turn are used to calculate rates.  The cross-sections themselves would change; however, cross-sections are simply intrinsic quantities that are used to calculate rates, so the cross-sections themselves would change by a certain factor, and the recipe for using cross-sections to calculate rates would change by the reciprocal of that factor, leaving the rates themselves entirely unchanged.  The non-relativistic limit of the cross-section would perhaps look a little less intuitive, but as a compensation we would no longer have to deal with ``relative velocities'' that exceed the speed of light.

\section*{Acknowledgments}

It is a pleasure to thank Ratin Akhoury for helpful discussions.

\end{document}